\documentclass[preprint,amsmath,amssymb,aps,showkeys,showpacs]{revtex4}
\usepackage[english]{babel}
\usepackage{graphicx}
\usepackage{graphics}
\usepackage{amsmath}
\usepackage{dcolumn}
\usepackage{amssymb}
\usepackage{bm}

\begin{document}
\title{On a neutral particle with a magnetic quadrupole moment in a uniform effective magnetic field}
\author{I. C. Fonseca}
\affiliation{Departamento de F\'isica, Universidade Federal da Para\'iba, Caixa Postal 5008, 58051-900, Jo\~ao Pessoa-PB, Brazil.}

\author{K. Bakke}
\email{kbakke@fisica.ufpb.br}
\affiliation{Departamento de F\'isica, Universidade Federal da Para\'iba, Caixa Postal 5008, 58051-900, Jo\~ao Pessoa-PB, Brazil.}

\begin{abstract}
Quantum effects on a Landau-type system associated with a moving atom with a magnetic quadrupole moment subject to confining potentials are analysed. It is shown that the spectrum of energy of the Landau-type system can be modified, where the degeneracy of the energy levels can be broken. In three particular cases, it is shown that the analogue of the cyclotron frequency is modified, and the possible values of this angular frequency of the system are determined by the quantum numbers associated with the radial modes and the angular momentum and by the parameters associated with confining potentials in order that bound states solutions can be achieved. 
\end{abstract}

\keywords{magnetic quadrupole moment, Landau quantization, hard-wall confining potential, linear confining potential, Coulomb-type confining potential, biconfluent Heun function, bound states}
\pacs{03.65.Ge, 31.30.jc, 31.30.J-, 03.65.Vf}

\maketitle

\section{Introduction}

It is well-known in the literature that the Landau quantization \cite{landau} takes place when the motion of a charged particle in a plane perpendicular to a uniform magnetic field acquires distinct orbits and the energy spectrum of this system becomes discrete and infinitely degenerate. It is important in studies of two-dimensional surfaces \cite{l2,l3,l4}, the quantum Hall effect \cite{l1} and Bose-Einstein condensation \cite{l5,l6}. With the aim of building a quantum system where the quantum Hall effect for neutral particles could be observed, it has been proposed in Ref. \cite{er} a model where the electric field that interacts with the permanent magnetic dipole moment of the neutral particle must satisfy specific conditions: the absence of torque on the magnetic dipole moment of the neutral particle, the electric field must satisfy the electrostatic conditions, and there exists the presence of a uniform effective magnetic field given by $\vec{B}_{\mathrm{eff}}=\vec{\nabla}\times\vec{A}_{\mathrm{eff}}$, where $\vec{A}_{\mathrm{eff}}=\vec{\sigma}\times\vec{E}$ corresponds to an effective vector potential, $\vec{E}$ is the electric field and $\vec{\sigma}$ are the Pauli matrices. Therefore, by choosing an electric field that satisfies the above conditions, it has been shown in Ref. \cite{er} that the motion of the neutral particle acquires discrete orbits, where the energy levels correspond to the analogue of the Landau levels. This analogue of the Landau levels corresponds to the Landau-Aharonov-Casher quantization \cite{er}. Another model of the Landau quantization for neutral particles has been proposed in Ref. \cite{lin} based on the dual effect of the Aharonov-Casher effect \cite{ac,hmw,hmw1}, where the neutral particle has a permanent electric dipole moment. In this model, there exists an effective vector potential given by $\vec{A}_{\mathrm{eff}}=\vec{\sigma}\times\vec{B}$, where $\vec{B}$ is a magnetic field that interacts with the permanent electric dipole moment of the neutral particle. This Landau-like quantization is called as the Landau-He-McKellar-Wilkens quantization \cite{lin}. Recently, the Landau quantization for an atom with electric quadrupole moment has been proposed in Ref. \cite{bf25} by imposing that the electric quadrupole tensor must be symmetric and traceless and there exists the presence of a uniform effective magnetic field given by an effective vector potential defined as $\vec{A}_{\mathrm{eff}}=\vec{Q}\times\vec{B}$, where $\vec{Q}$ is a vector associated with the electric quadrupole tensor \cite{chen,bf25} and $\vec{B}$ is the magnetic field in the laboratory frame.

The aim of this work is to analyse quantum effects on a Landau-type system associated with an atom with a magnetic quadrupole moment subject to some confining potentials. A great deal of work can be found in the literature with respect to studies of quadrupole moments of atoms and molecules, for instance, in single crystals \cite{quad}, refractive index \cite{quad2}, nuclear quadrupole interactions \cite{nucquad,nucquad2,quad4,quad14,quad16}, molecules \cite{quad-1,quad-2,quad3,quad15,quad17}, atoms \cite{quad6,prlquad}, superposition of chiral states \cite{quad18}, geometric quantum phases \cite{chen} and noncommutative quantum mechanics \cite{nonc}. In particular, quantum particles with a magnetic quadrupole moment have attracted interests in atomic systems \cite{magq5,magq6}, molecules \cite{magq7,magq9,magq11}, chiral anomaly \cite{magq}, with $P$- and $T$-odds effects in atoms \cite{magq3,magq8,magq10}, Coulomb-type interactions \cite{fb,fb4}, Landau-type quantization \cite{fb2,fb5} and analogue of the quantum Hall effect \cite{fb3}. Other interesting studies have been made in Refs. \cite{magq1,magq2,pra,prc,magq4}. In this work, we consider the single particle approximation used in Refs. \cite{prc,pra,fb,fb4} and then we deal with a system that consists in a moving atom with a magnetic quadrupole moment that interacts with external fields. Then, we introduce the Landau quantization associated with a moving atom that possesses a magnetic quadrupole moment, and thus we analyse the confinement of the Landau-type system to a hard-wall confining potential, a Coulomb-type potential, a linear confining potential and a Coulomb-type plus a linear confining potential.

The structure of this paper is as follows: in section II, we introduce the Landau-type system associated with an atom with a magnetic quadrupole moment by using the single particle approximation of Refs. \cite{prc,pra}; in section III, we confine the Landau-type system to a hard-wall confining potential and analyse the bound states solutions; in section IV, we discuss the Landau-type system subject to a Coulomb-type confining potential; in section V, we discuss the Landau-type system subject to a linear confining potential; in section VI, we discuss the Landau-type system subject to a Coulomb-type and a linear confining potentials; in section VII, we present our conclusions.

\section{Landau-type system}

In this section, we make a brief introduction of the Landau quantization associated with neutral particle (atom or molecule) with a magnetic quadrupole moment. First of all, by following Refs. \cite{pra,prc}, then, the single particle approximation that describes an atom with a magnetic quadrupole moment interacting with a magnetic field is given by a potential energy $U_{m}=-\sum_{i,j}M_{ij}\,\partial_{i}\,B_{j}$ in the rest frame of the particle, where $\vec{B}$ is the magnetic field and $M_{ij}$ is the magnetic quadrupole moment tensor, whose characteristic is that it is a symmetric and a traceless tensor. Recently, by considering a moving atom, it has been shown in Ref. \cite{fb} that the magnetic quadrupole moment of the atom interacts with a magnetic field given by $\vec{B}'=\vec{B}-\frac{1}{c^{2}}\,\vec{v}\times\vec{E}$ for $v\ll c$ \cite{griff}, where $c$ is the velocity of light. Thereby, the quantum dynamics of a moving atom with magnetic quadrupole moment can be described by the Schr\"odinger equation (with SI units)
\begin{eqnarray}
i\hbar\frac{\partial\psi}{\partial t}=\frac{1}{2m}\left[\hat{p}-\frac{1}{c^{2}}(\vec{M}\times\vec{E})\right]^2\,\psi-\vec{M}\cdot\vec{B}\,\psi,
\label{1.1}
\end{eqnarray} 
where vector $\vec{M}$ has the components determined by $M_{i}=\sum_{j}M_{ij}\,\partial_{j}$, the fields $\vec{E}$ and $\vec{B}$ given in Eq. (\ref{1.1}) are the electric and magnetic fields in the laboratory frame, respectively \cite{fb,griff}.

An analogue of the Landau quantization for a moving atom that possesses a magnetic quadrupole moment was proposed in Refs. \cite{fb2,fb3} based on the properties of the magnetic quadrupole tensor, and the field configuration in the laboratory frame that interacts with the magnetic quadrupole moment of the atom must produce a uniform effective magnetic field perpendicular to the plane of motion of the particle which is given by
\begin{eqnarray}
\vec{B}_{\mathrm{eff}}=\vec{\nabla}\times\left[\vec{M}\times\vec{E}\right],
\label{1.2}
\end{eqnarray}
 where the vector $\vec{E}$ is the the electric field in the laboratory frame and satisfies the electrostatic conditions. From this perspective, it has been shown in Ref. \cite{fb2} that an analogue of the Landau quantization can be obtained by considering the magnetic quadrupole moment tensor to be defined by the components:
\begin{eqnarray}
M_{\rho z}=M_{z\rho}=M,
\label{1.3}
\end{eqnarray}
where $M$ is a constant $\left(M>0\right)$ and with all other components of $M_{ij}$ as being zero, and thus there exists an electric field given by
\begin{eqnarray}
\vec{E}=\frac{\lambda\,\rho^{2}}{2}\,\hat{\rho},
\label{1.4}
\end{eqnarray}
where $\lambda$ is a constant associated with a non-uniform distribution of electric charges inside a non-conductor cylinder. In this particular case, we have that the magnetic quadrupole moment defined in Eq. (\ref{1.3}) is a symmetric and traceless matrix, and we also have an effective vector potential given by $\vec{A}_{\mathrm{eff}}=\vec{M}\times\vec{E}=\lambda\,M\,\rho\,\hat{\varphi}$ and, consequently, the effective magnetic field (\ref{1.2}) is uniform in the $z$-direction
\begin{eqnarray}
\vec{B}_{\mathrm{eff}}=\lambda\,M\,\hat{z},
\label{1.5}
\end{eqnarray}
that is, it is perpendicular to the plane of motion of the quantum particle. Therefore, the conditions for achieving the Landau quantization associated with a moving atom that possesses a magnetic quadrupole moment are satisfied. Henceforth, we are able to analyse the quantum effects of confining potentials on the Landau-type system described above.

\section{Landau-type system subject to a hard-wall confining potential}

In this section, we analyse the effects of the confinement of the Landau-type system established in the previous section to a hard-wall confining potential. By substituting Eqs. (\ref{1.3}) and (\ref{1.4}) into the Schr\"odinger equation (\ref{1.1}), we have (with the units $\hbar=c=1)$
\begin{eqnarray}
i\frac{\partial\psi}{\partial t}=-\frac{1}{2m}\left[\frac{\partial^{2}}{\partial\rho^{2}}+\frac{1}{\rho}\,\frac{\partial}{\partial\rho}+\frac{1}{\rho^{2}}\,\frac{\partial^{2}}{\partial\varphi^{2}}+\frac{\partial^{2}}{\partial z}\right]\psi-i\frac{M\,\lambda}{m}\,\frac{\partial\psi}{\partial\varphi}+\frac{M^{2}\,\lambda^{2}}{2m}\,\rho^{2}\,\psi.
\label{2.1}
\end{eqnarray}

A particular solution to Eq. (\ref{2.1}) is given in terms of the eigenvalues of the operators $\hat{p}_{z}=-i\partial_{z}$ and $\hat{L}_{z}=-i\partial_{\varphi}$ as 
\begin{eqnarray}
\psi\left(t,\,\rho,\,\varphi,\,z\right)=e^{-i\mathcal{E}t}\,e^{i\,l\,\varphi}\,e^{ikz}\,R\left(\rho\right),
\label{2.1a}
\end{eqnarray}
where $l=0,\pm1,\pm2,\ldots$ and $k$ is a constant, since these quantum operators  commute with the Hamiltonian of the right-hand side of Eq. (\ref{2.1}). By performing a change of variables $\xi=M\,\lambda\,\rho^{2}$ and, after some calculations, we have that the radial wave function is given by \cite{fb2}
\begin{eqnarray}
R\left(\xi\right)=\xi^{\frac{\left|l\right|}{2}}\,e^{-\frac{\xi}{2}}\,M\left(\frac{\left|l\right|}{2}+\frac{1}{2}-\frac{m\mathcal{E}}{2M\lambda}+\frac{l}{2},\,\left|l\right|+1,\,\xi\right),
\label{2.2}
\end{eqnarray}
 where $M\left(\frac{\left|l\right|}{2}+\frac{1}{2}-\frac{m\mathcal{E}}{2M\lambda},\,\left|l\right|+1,\,\xi\right)$ is the Kummer function of first kind or the confluent hypergeometric function \cite{abra}.

In condensed matter physics, a hard-wall confining potential is used with the purpose of describing a more realistic geometry of quantum dots and quantum rings as shown in Refs. \cite{mag,mag2,dot,dot2,fur,fur2}. For this purpose, let us now assume that the wave function of the atom is well-behaved at the origin, and thus vanishes at a fixed radius $\xi_{0}$, therefore we have 
\begin{eqnarray}
R\left(\xi_{0}=M\,\lambda\,\rho^{2}_{0}\right)=0.
\label{2.3}
\end{eqnarray}

Note that the parameter $\xi$ is defined in the range $0\,<\,\xi\,<\,\infty$ since the radial coordinate is defined in this range. However, by confining the system to a hard-wall confining potential, then, the parameter $\xi$ becomes defined in the range $0\,<\,\xi\,<\,\xi_{0}$. Therefore, we cannot impose that the confluent hypergeometric series becomes a polynomial of degree $n$ as in made in Ref. \cite{fb2} in order that the Landau quantization could be achieved because we now need to obtain a normalized wave function in the range $0\,<\,\xi\,<\,\xi_{0}$. A particular solution to Eq. (\ref{2.3}) is obtained by considering a fixed value for the parameter $b=\left|l\right|+1$ of the confluent hypergeometric function and the intensity of the electric field (\ref{1.4}) to be small, that is, we consider the parameter $\lambda$ in Eq. (\ref{1.4}) to be small. As a consequence, we can consider the product $M\lambda$ to be quite small, and hence the parameter $a=\frac{\left|l\right|}{2}+\frac{1}{2}-\frac{m\mathcal{E}}{2M\lambda}+\frac{l}{2}$ of the confluent hypergeometric function can be considered to be large, without loss of generality. Thereby, for a fixed $\xi_{0}$, the confluent hypergeometric function can be written in the form \cite{abra}:
\begin{eqnarray}
M\left(a,b,\xi_{0}=M\,\lambda\,\rho^{2}_{0}\right)&\approx&\frac{\Gamma\left(b\right)}{\sqrt{\pi}}\,e^{\frac{\xi_{0}}{2}}\left(\frac{b\xi_{0}}{2}-a\xi_{0}\right)^{\frac{1-b}{2}}\times\nonumber\\
[-2mm]\label{2.4}\\[-2mm]
&\times&\cos\left(\sqrt{2b\xi_{0}-4a\xi_{0}}-\frac{b\pi}{2}+\frac{\pi}{4}\right),\nonumber
\end{eqnarray}
where $\Gamma\left(b\right)$ is the gamma function. Next, by applying the boundary condition established in Eq. (\ref{2.3}), we obtain the following expression for the energy levels:
\begin{eqnarray}
\mathcal{E}_{n,\,l}\approx\frac{1}{2m\rho_{0}^{2}}\left[n\pi+\frac{\left|l\right|}{2}\pi+\frac{3\pi}{4}\right]^{2}+\frac{M\,\lambda\,l}{m}.
\label{2.5}
\end{eqnarray}

Thereby, Eq. (\ref{2.5}) is the spectrum of energy of the Landau-type system of an atom with a magnetic quadrupole moment subject to a hard-wall confining potential. Despite the presence of an effective uniform magnetic field characteristic of the Landau quantization, the influence of the hard-wall confining potential yields a spectrum of energy that differs from that of the analogue of the Landau quantization \cite{fb2}, where we have that the energy levels (\ref{2.5}) are parabolic with respect to the quantum number $n$ associated with radial modes and the analogue of the Landau levels are nonparabolic with respect to $n$.

\section{Landau-type system subject to a Coulomb-type confinement}

In this section, our focus is on the effects of a Coulomb-type interaction on the Landau-type system discussed in section II. Coulomb-type potentials have been reported as being in the interests of condensed matter physics, for instance, studies that have worked with 1-dimensional systems \cite{ct4,ct8,ct9,ct10,ct11}, molecules \cite{molecule,ct5,ct6}, position-dependent mass systems \cite{pdm2,pdm3,pdm5} and the Kratzer potential \cite{kratzer,kratzer2,kratzer3}. Furthermore, it is worth mentioning studies that have dealt with Coulomb-type potential in the propagation of gravitational waves \cite{ct14}, quark models \cite{quark} and relativistic quantum mechanics \cite{ct15,eug,cab3}. For this purpose, let us consider the Landau-type system to be subject to the confining potential:
\begin{eqnarray}
V\left(\rho\right)=\frac{\alpha}{\rho},
\label{6.1}
\end{eqnarray} 
where $\alpha$ is a constant that characterizes the Coulomb-type interaction. Since molecules can have a magnetic quadrupole moment \cite{magq7,magq9,magq11}, then, a particular interest in the Coulomb-type potential (\ref{6.1}) comes from the studies of molecular interactions described by the Kratzer potential \cite{kratzer,kratzer2,kratzer3}.

From Eqs. (\ref{2.1}) and (\ref{2.1a}), the radial equation with the Coulomb-type scalar potential (\ref{6.1}) is given by
\begin{eqnarray}
\left[2m\mathcal{E}+2\,M\,\lambda\,l\right]R=-R''-\frac{1}{\rho}\,R'+\frac{l^{2}}{\rho^{2}}\,R+M^{2}\,\lambda^{2}\,\rho^{2}\,R+\frac{2m\alpha}{\rho}\,R,
\label{6.2}
\end{eqnarray}
where we have also taken $k=0$. Now, we perform a change of variable given by $r=\sqrt{M\,\lambda}\,\,\rho$, and then, rewrite Eq. (\ref{6.2}) as
\begin{eqnarray}
R''+\frac{1}{r}\,R'-\frac{l^{2}}{\rho^{2}}\,R-r^{2}\,R-\frac{\nu}{r}\,R+\beta\,R=0,
\label{6.3}
\end{eqnarray}
where 
\begin{eqnarray}
\nu=\frac{2\,m\,\alpha}{\sqrt{M\,\lambda}};\,\,\,\,\,\,\beta=\frac{1}{M\,\lambda}\left[2m\mathcal{E}+2\,M\,\lambda\,l\right].
\label{6.4}
\end{eqnarray}

Observe that the asymptotic behaviour is determined for $r\rightarrow0$ and $r\rightarrow\infty$, then, we can write the function $R\left(r\right)$ in terms of an unknown function $G\left(r\right)$ as follows \cite{mhv,vercin,heun,fb}:
\begin{eqnarray}
R\left(r\right)=e^{-\frac{r^{2}}{2}}\,r^{\left|l\right|}\,G\left(r\right).
\label{6.5}
\end{eqnarray}

By substituting the function (\ref{6.5}) into Eq. (\ref{6.3}), we obtain the following equation for $G\left(r\right)$:
\begin{eqnarray}
G''+\left[\frac{2\left|l\right|+1}{r}-2r\right]\,G'+\left[\beta-2-2\left|l\right|-\frac{\nu}{r}\right]G=0,
 \label{6.6}
\end{eqnarray}
which is called as the biconfluent Heun equation \cite{heun}, and the function $G\left(r\right)$ is the biconfluent Heun function \cite{heun}: $G\left(r\right)=H_{B}\left(2\left|l\right|,\,0,\,\beta,\,2\nu,\,r\right)$. 

We proceed with using the Frobenius method \cite{arf}, then, we can write the solution to Eq. (\ref{6.6}) as a power series expansion around the origin: $G\left(r\right)=\sum_{k=0}^{\infty}a_{k}\,r^{k}$. By substituting this series into Eq. (\ref{6.6}), we obtain the recurrence relation:
\begin{eqnarray}
a_{k+2}=\frac{\nu}{\left(k+2\right)\left(k+2+2\left|l\right|\right)}\,a_{k+1}-\frac{\left(\beta-2-2\left|l\right|-2k\right)}{\left(k+2\right)\left(k+2+2\left|l\right|\right)}\,a_{k},
\label{6.7}
\end{eqnarray}

Let us start with $a_{0}=1$, then, from Eq. (\ref{6.7}), we can obtain other coefficients of the power series expansion, for instance, the coefficients $a_{1}$ and $a_{2}$:
\begin{eqnarray}
a_{1}&=&\frac{\nu}{\left(1+2\left|l\right|\right)};\nonumber\\
[-2mm]\label{6.8}\\[-2mm]
a_{2}&=&\frac{\nu^{2}}{2\,\left(2+2\left|l\right|\right)\left(1+2\left|l\right|\right)}-\frac{\left(\beta-2-2\left|l\right|\right)}{2\left(2+2\left|l\right|\right)}.\nonumber
\end{eqnarray}

By focusing on achieving bound states solutions, then, we need to impose that the biconfluent Heun series becomes a polynomial of degree $n$. From the recurrence relation (\ref{6.7}), we have that the biconfluent Heun series becomes a polynomial of degree $n$ by imposing that \cite{eug,fb}:
\begin{eqnarray}
\beta-2-2\left|l\right|=2n;\,\,\,\,\,a_{n+1}=0,
\label{6.9}
\end{eqnarray}
where $n=1,2,3,\ldots$. By analysing the first condition given in Eq. (\ref{6.9}), we obtain 
\begin{eqnarray}
\mathcal{E}_{n,\,l}=\varpi\left[n+\left|l\right|-l+1\right],
\label{6.10}
\end{eqnarray}
where $n=1,2,3,\ldots$ is the quantum number associated with the radial modes, $l=0,\pm1,\pm2,\ldots$ is the angular momentum quantum number and the angular frequency of the system becomes
\begin{eqnarray}
\varpi=\frac{M\,\lambda}{m}.
\label{6.10a}
\end{eqnarray}
Note that the the angular frequency is modified in contrast to the cyclotron frequency $\omega=\frac{2M\lambda}{m}$ obtained in Ref. \cite{fb2}.

Next, we analyse the condition $a_{n+1}=0$ given in Eq. (\ref{6.9}) for the ground state of the system ($n=1$). Let us assume that the angular frequency $\varpi$ can be adjusted in such a way that the condition $a_{n+1}=0$, therefore, we shall label $\varpi=\varpi_{n,\,l}$ from now on. As a result of this assumption, we have that both conditions imposed in Eq. (\ref{6.9}) are satisfied and a polynomial solution to the function $G\left(r\right)$ is obtained. Hence, for $n=1$ we have that $a_{n+1}=a_{2}=0$, and thus the possible values of the angular frequency $\varpi$ associated with the ground state of the system are given by
\begin{eqnarray}
\varpi_{1,\,l}=\frac{2m\alpha^{2}}{\left(1+2\left|l\right|\right)},
\label{6.11}
\end{eqnarray}
which shows us that only some specific values of the angular frequency of the harmonic oscillator are allowed in the system in order that bound state solutions can be obtained. Besides, from Eq. (\ref{6.11}), we have that the possible values of the angular frequency are determined by the quantum numbers $\left\{n,\,l\right\}$ of the system and the parameter associated with Coulomb-type interaction. By substituting (\ref{6.11}) into Eq. (\ref{6.10}), we have that the energy of the ground state of the system is given by
\begin{eqnarray}
\mathcal{E}_{1,\,l}=\frac{2m\alpha^{2}}{\left(1+2\left|l\right|\right)}\,\left[\left|l\right|-l+2\right].
\label{6.12}
\end{eqnarray}

Hence, the general expression for the energy levels (\ref{6.10}) can be written as
\begin{eqnarray}
\mathcal{E}_{n,\,l}=\varpi_{n,\,l}\left[n+\left|l\right|-l+1\right].
\label{6.13}
\end{eqnarray}

In contrast to Ref. \cite{fb2}, we have that the energy levels of the Landau-type system are modified due to the influence of the Coulomb-type interaction. Moreover, the angular frequency is also modified by the effects of the Coulomb-type interaction, where the possible values of the angular frequency $\varpi$ are determined by the quantum numbers of the system $\left\{n,\,l\right\}$ and the parameter associated with Coulomb-type interaction. The ground state of the system becomes determined by the quantum number $n=1$ instead of the quantum number $n=0$ given in Ref. \cite{fb2}, and the degeneracy of the analogue of the Landau levels is broken.

\section{Landau-type system subject to a linear confining potential}

Several works have dealt with a linear scalar potential in molecular and atomic physics through the perturbation theory \cite{linear3a,linear3b,linear3c,linear3d,linear3e,linear3f}. In the context of the quantum field theory, the interest in the linear scalar potential comes from the studies of confinement of quarks \cite{linear1a,linear1b,linear4a,linear4b,linear4c,linear4d}. Besides, relativistic quantum systems \cite{,linear2b,linear2c,linear2d,linear2e,linear2f,linear2g,eug,scalar,scalar2,vercin,mhv} have shown a great interest in the linear scalar potential. Now, we focus on the effects of a linear confining potential on the Landau-type system for an atom/molecule with a magnetic quadrupole moment by searching for analytical solutions. In this way, we introduce the following scalar potential into the Schr\"odinger equation (\ref{2.1}):
\begin{eqnarray}
V\left(\rho\right)=\eta\,\rho,
\label{7.1}
\end{eqnarray} 
where $\eta$ is a constant that characterizes the linear interaction. Then, by following the steps from Eq. (\ref{2.1}) to Eq. (\ref{2.1a}), then, the radial equation with the linear confining potential (\ref{7.1}) becomes 
\begin{eqnarray}
\left[2m\mathcal{E}+2\,M\,\lambda\,l\right]R=-R''-\frac{1}{\rho}\,R'+\frac{l^{2}}{\rho^{2}}\,R+M^{2}\,\lambda^{2}\,\rho^{2}\,R+2m\,\eta\,\rho\,R.
\label{7.2}
\end{eqnarray}
Let us also take $k=0$ and perform the same change of variable made in the previous section, i.e., $r=\sqrt{M\,\lambda}\,\,\rho$; thus, we have
\begin{eqnarray}
R''+\frac{1}{r}\,R'-\frac{l^{2}}{\rho^{2}}\,R-r^{2}\,R-\theta\,r\,R+\beta\,R=0,
\label{7.3}
\end{eqnarray}
where the parameter $\beta$ has been defined in Eq. (\ref{6.4}) and the parameter $\theta$ is defined as
\begin{eqnarray}
\theta=\frac{2\,m\,\eta}{\left(M\,\lambda\right)^{3/2}}.
\label{7.4}
\end{eqnarray}

By analysing the asymptotic behaviour as in the previous section, we can write the function $R\left(r\right)$ in terms of an unknown function $H\left(r\right)$ as follows \cite{mhv,vercin,heun,fb}:
\begin{eqnarray}
R\left(r\right)=e^{-\frac{r^{2}}{2}}\,e^{-\frac{\theta\,r}{2}}\,r^{\left|l\right|}\,H\left(r\right).
\label{7.5}
\end{eqnarray}

By substituting the function (\ref{7.5}) into Eq. (\ref{7.3}), we obtain the following equation for $H\left(r\right)$:
\begin{eqnarray}
H''+\left[\frac{2\left|l\right|+1}{r}-\theta-2r\right]\,H'+\left[\beta+\frac{\theta^{2}}{4}-2-2\left|l\right|-\frac{\theta\left(2\left|l\right|+1\right)}{2r}\right]H=0,
 \label{7.6}
\end{eqnarray}
which is the biconfluent Heun equation \cite{heun}, and the function $H\left(r\right)$ is the biconfluent Heun function \cite{heun}: $H\left(r\right)=H_{B}\left(2\left|l\right|,\,\theta,\,\beta+\frac{\theta^{2}}{4},\,0,\,r\right)$. Further, by following the steps from Eq. (\ref{6.7}) to Eq. (\ref{6.8}), we obtain a recurrence relation given by
\begin{eqnarray}
a_{k+2}=\frac{\theta\left(2k+3+2\left|l\right|\right)}{2\left(k+2\right)\left(k+2+2\left|l\right|\right)}\,a_{k+1}-\frac{\left(4\beta+\theta^{2}-8-8\left|l\right|-8k\right)}{4\left(k+2\right)\left(k+2+2\left|l\right|\right)}\,a_{k},
\label{7.7}
\end{eqnarray}
and with $a_{0}=1$, we have that the coefficients $a_{1}$ and $a_{2}$ are given by
\begin{eqnarray}
a_{1}&=&\frac{\theta}{2};\nonumber\\
[-2mm]\label{7.8}\\[-2mm]
a_{2}&=&\frac{\theta^{2}\left(2\left|l\right|+3\right)}{8\,\left(2+2\left|l\right|\right)}-\frac{\left(4\beta+\theta^{2}-8-8\left|l\right|\right)}{8\left(2+2\left|l\right|\right)}.\nonumber
\end{eqnarray}

By imposing that the biconfluent Heun series becomes a polynomial of degree $n$, then, we have from Eq. (\ref{7.7}), we have that the biconfluent Heun series becomes a polynomial of degree $n$ when:
\begin{eqnarray}
4\beta+\theta^{2}-8-8\left|l\right|=8n;\,\,\,\,\,a_{n+1}=0,
\label{7.9}
\end{eqnarray}
where $n=1,2,3,\ldots$. The first condition given in Eq. (\ref{7.9}) yields 
\begin{eqnarray}
\mathcal{E}_{n,\,l}=\varpi\left[n+\left|l\right|-l+1\right]-\frac{\eta^{2}}{2m\varpi^{2}},
\label{7.10}
\end{eqnarray}
where $n=1,2,3,\ldots$ is the quantum number associated with the radial modes, $l=0,\pm1,\pm2,\ldots$ is the angular momentum quantum number and $\varpi$ is the angular frequency of the system given in Eq. (\ref{6.10a}).

Further, let us analyse the second condition $a_{n+1}=0$ given in Eq. (\ref{7.9}) by considering the ground state of the system as in the previous section. Again, we assume that the angular frequency $\varpi$ is a parameter that can be adjusted in order to satisfy the condition $a_{n+1}=0$. For $n=1$, we obtain  
\begin{eqnarray}
\varpi_{1,\,l}=\left[\frac{\eta^{2}}{2m}\left(2\left|l\right|+3\right)\right]^{1/3},
\label{7.11}
\end{eqnarray}
where we also have that the possible values of the angular frequency are determined by the quantum numbers of the system $\left\{n,\,l\right\}$ and, in this case, by the parameter associated with the linear interaction. Besides, from Eqs. (\ref{7.10}) and (\ref{7.11}), the energy level associated with the ground state is
\begin{eqnarray}
\mathcal{E}_{1,\,l}=\left[\frac{\eta^{2}}{2m}\left(2\left|l\right|+3\right)\right]^{1/3}\times\left[\left|l\right|-l+2\right]-\frac{\eta^{2}}{2m}\left(\frac{2m}{\eta^{2}\left[2\left|l\right|+3\right]}\right)^{2/3}.
\label{7.12}
\end{eqnarray}

Hence, the general form the energy levels of the Landau-type system under the influence of a linear confining potential can be written as 
\begin{eqnarray}
\mathcal{E}_{n,\,l,\,s}=\varpi_{n,\,l}\left[n+\left|l\right|-l+1\right]-\frac{\eta^{2}}{2m\,\varpi_{n,\,l}^{2}}.
\label{7.13}
\end{eqnarray}

By comparing the spectrum of energy (\ref{7.13}) with the analogue of the Landau levels, we have that the energy levels are modified by the linear confining potential. The effects of the linear interaction yields a new contribution to the energy levels given by the last term of Eqs. (\ref{7.10}) and (\ref{7.13}). The angular frequency is also modified due to the linear interaction in contrast to the cyclotron frequency $\omega=\frac{2M\lambda}{m}$ obtained in Ref. \cite{fb2}, but it has the same form of that obtained in the previous section for the Coulomb-type interaction case given in Eq. (\ref{6.10a}). Furthermore, in this case, the possible values of the angular frequency $\varpi$ are determined by the quantum numbers of the system $\left\{n,\,l\right\}$ and by the parameter associated with the linear interaction. The ground state of the system also becomes determined by the quantum number $n=1$ instead of the quantum number $n=0$ of the Landau-like levels. Finally, from Eq. (\ref{7.10}) to Eq. (\ref{7.13}), we have that the degeneracy of the analogue of the Landau levels is broken.

\section{Landau-type system subject to a Coulomb-type plus linear confining potential}

Linear plus Coulomb-type potential has been studied in the context the perturbation theory \cite{linear3g}, with the WKB approximation \cite{linear5a} and the Bohr-Sommerfeld quantization \cite{linear5b}. Analytical solutions to the Schr\"odinger equation in the presence of a linear and Coulomb-type potentials have been discussed in the context of the quark systems \cite{linear2a} and in mesoscopic systems. In this section, we consider the Landau-type system for an atom/molecule with a magnetic quadrupole moment to be subject to the Coulomb-type plus linear confining potential and search for analytical solutions. Thereby, let us introduce a scalar potential into the Schr\"odinger equation (\ref{2.1}) given by
\begin{eqnarray}
V\left(\rho\right)=\frac{\alpha}{\rho}+\eta\,\rho,
\label{8.1}
\end{eqnarray}
where $\alpha$ and $\eta$ are constants as we have established previously. From Eqs. (\ref{2.1}) and (\ref{2.1a}), the radial equation becomes
\begin{eqnarray}
R''+\frac{1}{r}\,R'-\frac{l^{2}}{\rho^{2}}\,R-r^{2}\,R-\theta\,r\,R-\frac{\nu}{r}\,R+\beta\,R=0,
\label{8.2}
\end{eqnarray}
where that parameters $\nu$, $\beta$ and $\theta$ have been defined in Eqs. (\ref{6.4}) and (\ref{7.4}), respectively. By analysing the asymptotic behaviour, we can write the radial wave function $R\left(r\right)$ in the same way as given in Eq. (\ref{7.5}); thus, substituting the radial wave function (\ref{7.5}) into Eq. (\ref{8.2}), we obtain the following second order differential equation
By substituting the function (\ref{7.5}) into Eq. (\ref{7.3}), we obtain the following equation for $H\left(r\right)$:
\begin{eqnarray}
H''+\left[\frac{2\left|l\right|+1}{r}-\theta-2r\right]\,H'+\left[\beta+\frac{\theta^{2}}{4}-2-2\left|l\right|-\frac{\theta\left(2\left|l\right|+1\right)+2\nu}{2r}\right]H=0,
 \label{8.3}
\end{eqnarray}
which is also the biconfluent Heun equation \cite{heun}, and the function $H\left(r\right)$ is the biconfluent Heun function \cite{heun}: $H\left(r\right)=H_{B}\left(2\left|l\right|,\,\theta,\,\beta+\frac{\theta^{2}}{4},\,2\nu,\,r\right)$. Further, by following the steps from Eq. (\ref{6.7}) to Eq. (\ref{6.8}), we obtain a recurrence relation given by
\begin{eqnarray}
a_{k+2}=\frac{\theta\left(2k+3+2\left|l\right|\right)+2\nu}{2\left(k+2\right)\left(k+2+2\left|l\right|\right)}\,a_{k+1}-\frac{\left(4\beta+\theta^{2}-8-8\left|l\right|-8k\right)}{4\left(k+2\right)\left(k+2+2\left|l\right|\right)}\,a_{k},
\label{8.4}
\end{eqnarray}
and with $a_{0}=1$, we have that the coefficients $a_{1}$ and $a_{2}$ are given by
\begin{eqnarray}
a_{1}&=&\frac{\theta}{2}+\frac{\nu}{\left(1+2\left|l\right|\right)};\nonumber\\
[-2mm]\label{8.5}\\[-2mm]
a_{2}&=&\frac{\theta^{2}\left(3+2\left|l\right|\right)}{8\left(2+2\left|l\right|\right)}+\frac{\theta\,\nu\left(1+\left|l\right|\right)}{\left(2+2\left|l\right|\right)\left(1+2\left|l\right|\right)}+\frac{\nu^{2}}{2\left(2+2\left|l\right|\right)\left(1+2\left|l\right|\right)}\nonumber\\
&-&\frac{\left(4\beta+\theta^{2}-8-8\left|l\right|\right)}{8\left(2+2\left|l\right|\right)}.\nonumber
\end{eqnarray}

Again, we must impose that the biconfluent Heun series becomes a polynomial of degree $n$, then, from Eq. (\ref{8.4}), we have that the biconfluent Heun series becomes a polynomial of degree $n$ with the following conditions:
\begin{eqnarray}
4\beta+\theta^{2}-8-8\left|l\right|=8n;\,\,\,\,\,a_{n+1}=0,
\label{8.6}
\end{eqnarray}
where $n=1,2,3,\ldots$. The first condition of Eq. (\ref{8.6}) yields a general expression for the energy levels:
\begin{eqnarray}
\mathcal{E}_{n,\,l}=\varpi\left[n+\left|l\right|-l+1\right]-\frac{\eta^{2}}{2m\varpi^{2}},
\label{8.7}
\end{eqnarray}
where $n=1,2,3,\ldots$ is the quantum number associated with the radial modes, $l=0,\pm1,\pm2,\ldots$ is the angular momentum quantum number and $\varpi$ is the angular frequency of the system given in Eq. (\ref{6.10a}).

From the second condition $a_{n+1}=0$ given in Eq. (\ref{8.6}), where we also assume that the angular frequency $\varpi$ is a parameter that can be adjusted in order to satisfy the condition $a_{n+1}=0$, we have for the ground state of the system ($n=1$) that the possible values of the angular frequency associated with the ground state is determined by the following third-degree algebraic equation \cite{eug,fb4}:
\begin{eqnarray}
\varpi^{3}_{1,\,l}-\frac{2m\alpha^{2}}{\left(1+2\left|l\right|\right)}\,\varpi^{2}_{1,\,l}-\frac{4\left(1+\left|l\right|\right)}{\left(1+2\left|l\right|\right)}\,\varpi_{1,\,l}-\frac{\eta^{2}}{2m}\,\left(3+2\left|l\right|\right)=0.
\label{8.8}
\end{eqnarray}
Despite Eq. (\ref{8.8}) has at least one real solution, we do not write it because its expression is very long. Besides, the general expression for the energy levels (\ref{8.7}) should be written in the same form of Eq. (\ref{7.13}).

Hence, we have obtained analytical solutions to the Landau-type system for an atom/molecule with a magnetic quadrupole moment to be subject to the Coulomb-type and linear confining potentials. Note that the possible values of the angular frequency $\varpi$ are determined by the quantum numbers of the system $\left\{n,\,l\right\}$ and by the parameters associated with the linear and Coulomb-type confining potentials in order that a polynomial solution to the function $H\left(r\right)$ can be obtained. The ground state of the system is also determined by the quantum number $n=1$ instead of the quantum number $n=0$ of the Landau-like levels. Finally, from Eq. (\ref{8.7}) to Eq. (\ref{8.8}), we have that the degeneracy of the analogue of the Landau levels is broken.

\section{conclusions}

We have investigated the behaviour of a neutral particle (atom or molecule) with a magnetic quadrupole moment in a region with a uniform effective magnetic field subject to confining potentials. We have analysed the confinement of the Landau-type system to a hard-wall confining potential, a Coulomb-type potential, a linear confining potential and a linear plus Coulomb-type potential. In the confinement to a hard-wall confining potential, we have seen that the spectrum of energy is modified in contrast to the Landau-type levels, where the energy levels are parabolic with respect to the quantum number associated with the radial modes. 

On the other hand, with respect to the confinement to a Coulomb-type potential, a linear confining potential and a linear plus Coulomb-type potential, we have obtained different spectrum of energies. In these three cases, the ground state of the system becomes determined by the quantum number $n=1$ instead of the quantum number $n=0$ obtained in the Landau-like levels, and the degeneracy of the analogue of the Landau levels is broken. Moreover, the cyclotron frequency of the Landau-type system is modified by the influence of the confining potentials, where the possible values of the angular frequency of the system are determined by the quantum numbers $\left\{n,\,l\right\}$ and by the parameters associated with confining potentials. In particular, in the case of the confinement to the linear plus Coulomb-type potential, the possible values of the angular frequency associated with the ground state is determined by a third-degree algebraic equation.

\acknowledgments

The authors would like to thank the Brazilian agencies CNPq and CAPES for financial support.

\end{document}